\documentclass[12pt]{iopart}
\usepackage{graphicx}
\usepackage{amssymb}
\newcommand{\VmA}{V\!\!\!-\!\!A}
\begin{document}

\title{Lattice QCD for Precision Nucleon Matrix Elements}

\author{Huey-Wen Lin\footnote{NT@UW-11-31}}

\address{Department of Physics, University of Washington, Seattle, WA 98195-1560}
\ead{hwlin@phys.washington.edu}

\begin{abstract}
Precision measurements on nucleons provide constraints on the Standard Model and can also discern the signatures predicted for particles beyond the Standard Model. Knowing the Standard Model inputs to nucleon matrix elements will be necessary to constrain the couplings of dark matter candidates such as the neutralino, to relate the neutron electric dipole moment to the CP-violating theta parameter, or to search for new TeV-scale particles though non-$\VmA$ interactions in neutron beta decay. However, these matrix elements derive from the properties of quantum chromodynamics (QCD) at low energies, where the coupling is strong and thus perturbative treatments fail. Using lattice gauge theory, we can nonperturbatively calculate the QCD path integral on a supercomputer.

In this proceeding, I will review a few representative areas in which lattice QCD can contribute to understanding the structure inside nucleon and how they can contribute to the search for beyond-the-Standard Model physics, with discussions of the difficulties and prospects for future development.
\end{abstract}

\maketitle

\section{Introduction}
Quantum chromodynamics (QCD) is the theory of the physics that dominates at very small scales, the femtoscale ($10^{-15}$~m), where fundamental particles (quarks and gluons) are bound inside the building blocks of matter, protons and neutrons. This is far smaller than the size of the atoms ($10^{-10}$~m) which form molecules (nanoscale, $10^{-9}$) in the materials we are familiar with, like a cup of coffee.
What do we learn by considering physics at this tiny scale?
For one, we can understand how fundamental particles make up the proton and neutron, and more exotic particles such as hyperons. Further, we may try to see how they interact with each other.
At the nuclear level, we can learn how the residual strong force binds nucleons together into the nuclei that form the cores of all atoms, that power the stars and perhaps one day fusion power plants. An understanding of how nuclear structure arises from fundamental physics may hold the key to the question of fine tuning (a mystery relevant to any carbon-based life forms).
The interactions between hadrons at small scale also impact objects at the astrophysical scale, such as neutron stars. The strengths of these interactions can be input into descriptions of nuclear matter in simulations of the conditions in the center of neutron stars,
where incredible pressure and density could allow exotic forms of matter to exist.

\begin{figure}[t]
\begin{center}
\includegraphics[width=0.75\textwidth]{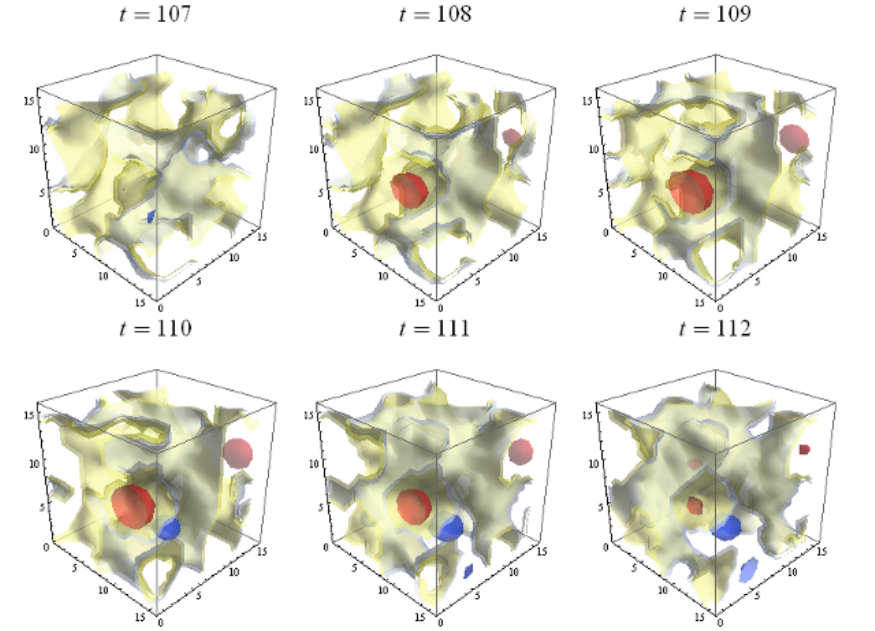}
\end{center}
\caption{\label{fig:instantons}
A selection of timeslices showing the topological charge of the vacuum calculated on one gauge-field configuration on a lattice with spacing approximately $0.1227 \times 10^{-15}$ meters with 16 lattice sites in each spatial direction. The evolution of the vacuum shown in this image occurs over $5.8 \times 10^{-25}$ seconds. The red regions correspond to instantons with positive topological winding number, while the blue correspond to instantons with negative winding number.
}
\end{figure}

Quantum chromodynamics is the theory of the color force. It describes the strong interactions between quarks and gluons using an SU(3) gauge theory. Given a QCD action $S$ and interesting observables whose properties are described by an operator ${\cal O}$, one can compute the physical quantities of interest using a path integral (integrating over all possible configurations of gluonic and fermionic fields $A$, $\psi$, $\bar{\psi}$):
\begin{eqnarray}
\langle {\cal O}[A,\psi,\bar{\psi}] \rangle = 
     \frac{1}{Z} \int DA\,D\psi\,D\bar{\psi}\;
                       e^{iS[A,\psi,\bar{\psi}]} {\cal O}[A,\psi,\bar{\psi}].
\end{eqnarray}
One of the interesting properties of quantum chromodynamics is confinement. That is, we never see free quarks in nature; rather, all we see are composite particles containing quarks, called hadrons, such as the lightest hadron, the pion, and the proton and neutron of ordinary nuclei.
At large energy, the QCD coupling is small. We can simply make an expansion in terms of the coupling, and it converges well. In fact, the calculations can be easy enough (at low order) to give to your graduate field-theory class as homework.
However, at low energy the strong coupling becomes larger and larger, and, by say the scale of neutron beta decay, perturbation theory becomes poorly convergent.
At low energy, the theory becomes nonperturbative and even just the vacuum of QCD is incredibly complicated.
The QCD vacuum teems with topological charge and the results of the broken quark chiral symmetry.
Figure~\ref{fig:instantons} shows a few timeslices of the spatial distribution of topological charge in an example QCD vacuum configuration. 
Unlike the classical vacuum, just to describe this is incredibly complicated and a task that is nearly impossible analytically.
Therefore, nonperturbative approaches are essential to exploring QCD physics at this energy scale.
In order to study the interesting physics in the low-energy regions, in the 1970s Kenneth Wilson proposed to discretize space and time in the path integral and to work in Euclidean space. Thus, lattice QCD (LQCD) was born. It contains two scales that are absent in continuum QCD, one ultraviolet (the lattice spacing $a$) and one infrared (the spatial extent of the box $L$). The problem now only involves a finite number of degrees of freedom and can be put on a computer for numerical integration.

Of course, if computational resources were not an issue, we could solve all the QCD problems by brute force as long as we could write down the path integrals.
However, it is not possible within a human lifetime to do so, so we have to examine the problem and solve it more cleverly.
The number of degrees of freedom for a fermion here is the spatial volume times the number (3) of colors, times the number (4) of spins. The number of spatial points varies, for example, a lattice volume with $64^3\times 96$ has rank 300M, and it cannot be fit onto a few workstations. A computational cluster with thousands of cores is necessary. Lattice practitioners use systems ranging from small-size clusters hosted at universities to national supercomputer centers, such as Kraken at NICS, or CPU-GPU hybrid systems, such as Edge at LLNL.

\begin{figure}
\begin{center}
\includegraphics[height=.23\textheight]{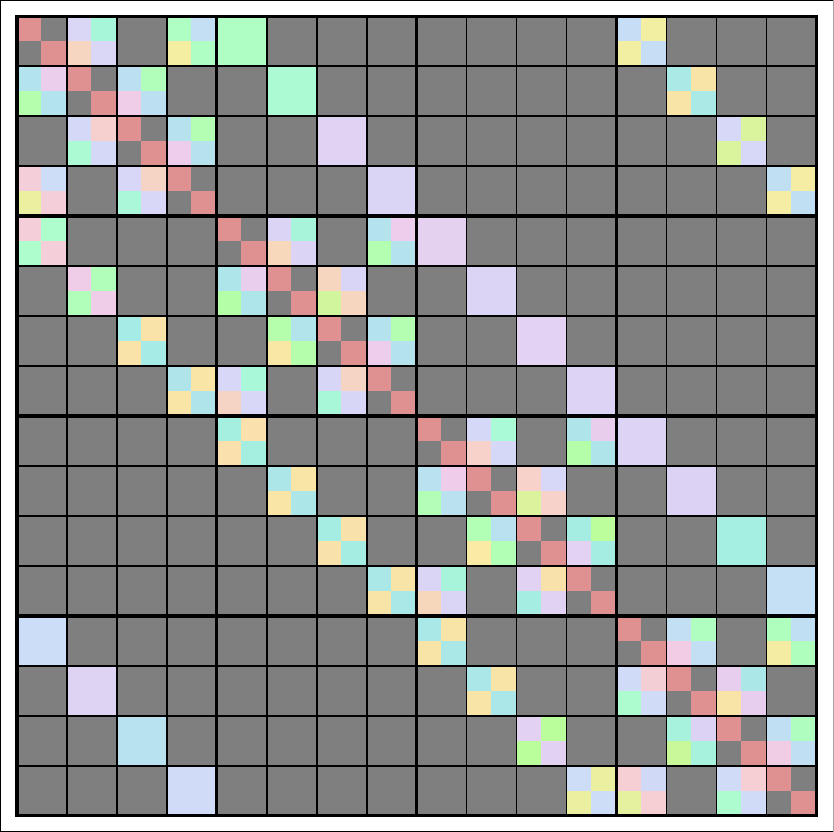}
\includegraphics[height=.23\textheight]{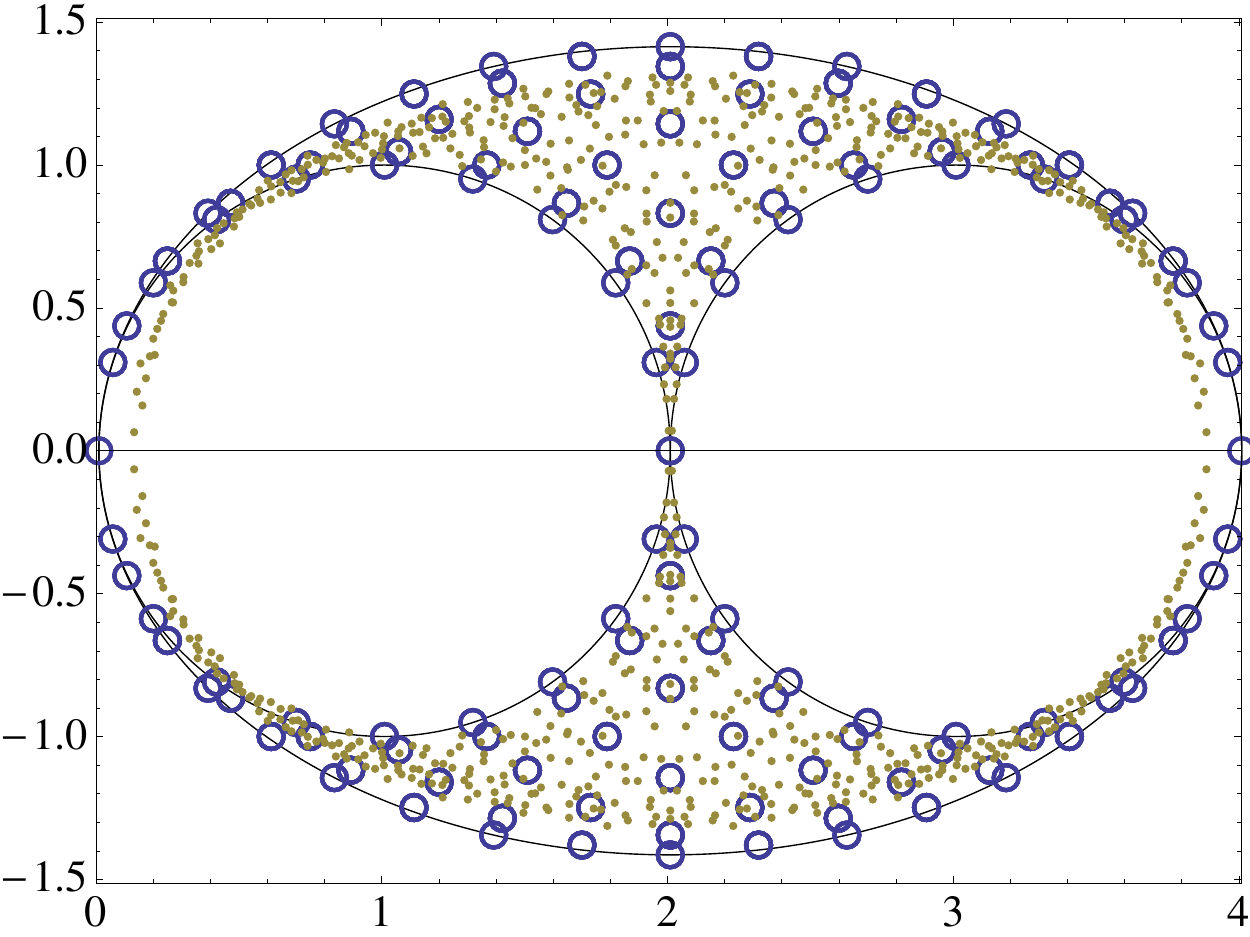}
\caption{\label{fig:2D-Wilson} (left) Two-dimensional illustration of the spin structure of the Wilson Dirac operator. Gray is zero; blue is real and positive; red is negative; magenta and green are positive and negative imaginary.
(right) Spectral plot for a two-dimensional Wilson Dirac operator. The blue circles indicate the results from free-field, while the olive dots have U(1) interacting gauge fields. The smaller the $m_q$, the worse the matrix condition number.
}
\end{center}
\end{figure}

In most cases, we spend the majority of computational hours inverting the Dirac operator. Due to the breaking of continuous rotation symmetry, multiple versions of the Dirac operator can be written on the lattice that describe the same physics in the continuum.
Such a discrete Dirac operator is a sparse and structured matrix, properties that we can take advantage of to develop smarter solvers for our problem.
To illustrate structure of our operator, we show a 2-dimensional version with U(1) gauge fields (in reality, we use a much more complicated 4D, SU(3) theory); this particular form is the Wilson Dirac operator:
\begin{equation}
\label{eq:WilsonD}
D^{\rm Wilson}_{x,x^\prime} =
(m+d)\,\delta_{x,x^\prime}
- \sum_{\mu} \frac{1}{2}
     \left[ (1+\gamma_\mu) U_{x,\mu} \delta_{x+\hat{\mu},x^\prime} +
        (1-\gamma_\mu) U_{x,\mu}^\dag \delta_{x,x^\prime+\hat{\mu}} \right].
\end{equation}
The left-hand side of Fig.~\ref{fig:2D-Wilson} shows the spin structure of the two-dimensional version of Eq.~\ref{eq:WilsonD}. We can see that it is highly structured along diagonals, with non-zero terms in regular places, but mostly zero.
The eigenvalues of this operator in the complex plane are shown on the right-hand side of Fig.~\ref{fig:2D-Wilson}. When interactions are turned off, the eigenvalues are these blue circles, while the olive points are the eigenvalues with thermalized U(1) gauge fields included.
As the quark mass is decreased, entire eigenspectrum is shifted to the left; thus, the lowest eigenvalues approach zero and the matrix condition number diverges. This means that our inversion algorithms will run very slowly.
Only recently have ensembles with physical pion masses become available; most of the time, we do the calculation with multiple heavier masses, and a procedure for extrapolation is necessary.
Taking advantage of the structure of the operator, many techniques have been developed to solve the Dirac equation faster:
eigenvector deflation, multi-mass solvers, multigrid acceleration, stochastic sources.
Some such techniques have been examining the memory/communication problem, which will be needed to run QCD effectively on systems including GPUs~\cite{Joo:2011xx}.

Once we have $D^{-1}$, we can use it to calculate Green functions from which we will extract physical quantities.
For example, we can calculate the pion mass by creating a pion at a certain point in space and time and annihilating it at another. In practice, we ``contract'' the inverse Dirac matrix with some spin/color matrix:
\begin{eqnarray}
\label{eq:pion-prop}
C_\pi(t) = \left\langle\pi(0)|\pi(t)\right\rangle 
= \left\langle \overline{u}\gamma_5d(0)\,\,\overline{d}\gamma_5u(t) \right\rangle 
= \left\langle D^{-1}(0,t)(D^{-1}(0,t))^\dag \right\rangle .
\end{eqnarray}
After momentum projection to the rest frame, the resulting Green function declines exponentially in the time direction with exponent equal to the mass.

\section{Probing inside Nucleons}

To ``see'' inside a hadron, we need a stronger ``microscope'' than the typical optical ones; to probe inside the proton requires, for example, a GeV-scale electron beam elastically scattering with a nucleon.
This creates interaction with the nucleons through exchange of photons; in QCD, it can be simulated as a vector current to probe electromagnetic properties. We still create the particle and annihilate it at another coordinate of spacetime;
in addition, we now insert a vector current to probe the nucleon's structure in the following general form:
\begin{eqnarray}
\label{eq:threepoint}
\Gamma^{(3)} = \langle N(x_2) J(x_1) \overline{N}(x_0) \rangle,
\end{eqnarray}
where
\begin{eqnarray}
J(x) = \sum_f e_f \overline{q}_f(x){\cal O}(x,y) q_f(y),
\end{eqnarray}
and we sum over all flavors $f \in \{u,d\}$; $\cal O$ can be either simple gamma matrices or their products with some covariant displacement function. ${\cal O}=\gamma_\mu \delta_{x,y}$ is used for simulating the elastic scattering process.
Such a current can connect directly to the quark fields in the baryon operator or it can interact with vacuum quark loops. The available momenta in the system are limited by the lattice's boundary conditions, the box size and lattice spacing. By varying the inserted operator $J$, we can study various properties of nucleons. In this section, we will present a few examples on nucleon axial charge, proton spin and neutron transverse density.

\subsection{Nucleon Axial Charge $g_A$}
\begin{figure}
\begin{center}
\includegraphics[height=.21\textheight]{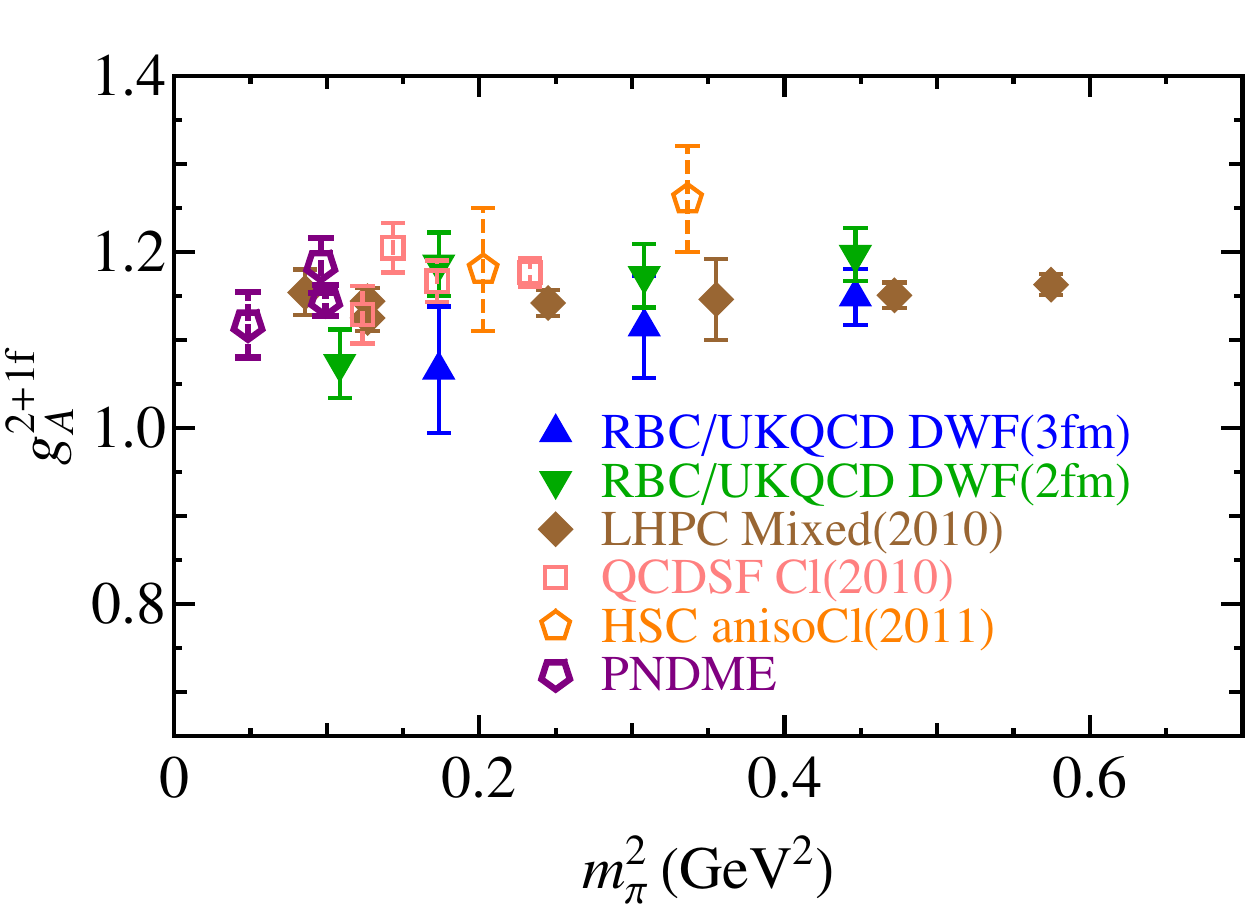}
\includegraphics[height=.2\textheight]{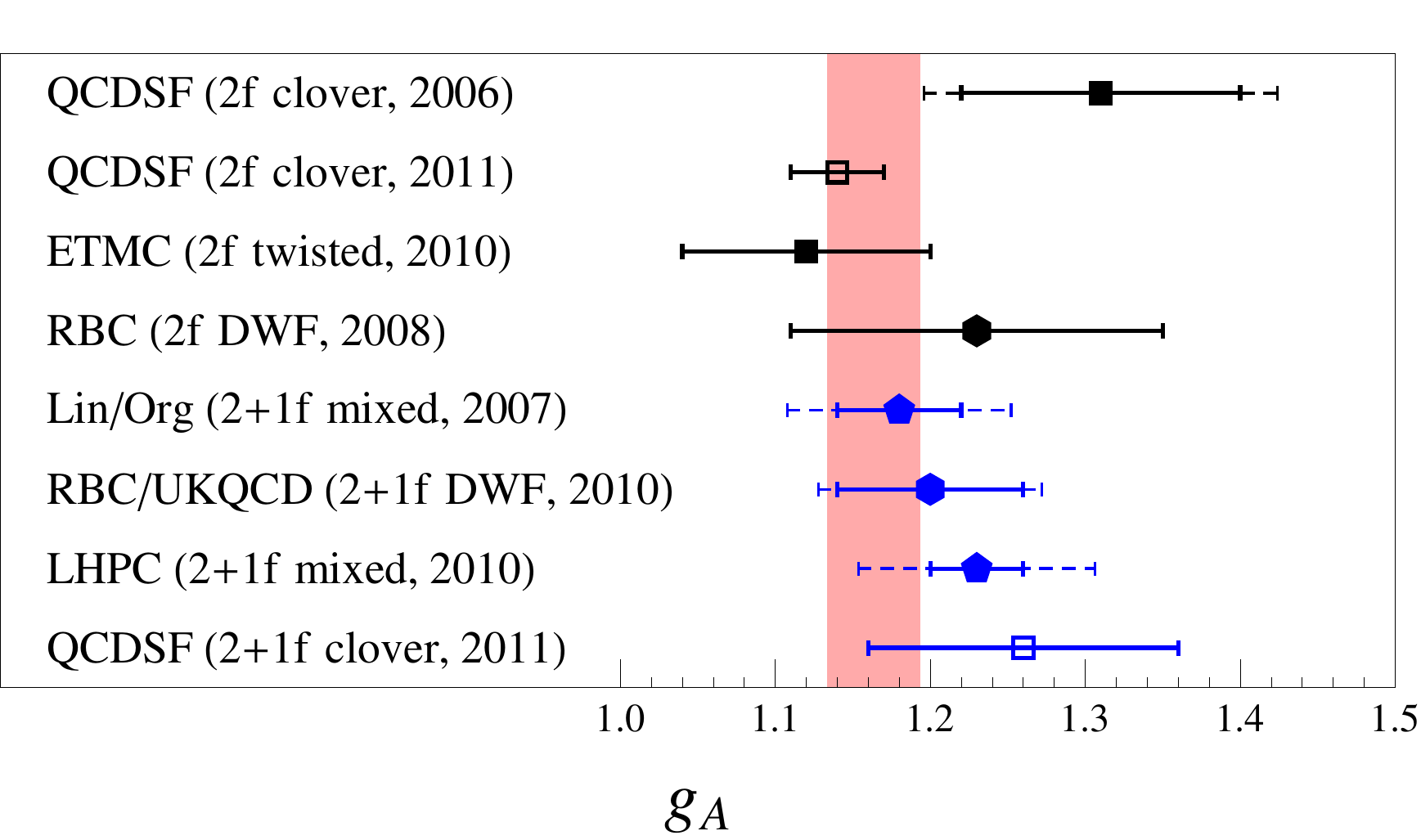}
\end{center}
\caption{\label{fig:gAT3f} (left) Summary of the lattice $N_f=2+1$ results on $g_A$ as functions of $m_\pi^2$~\cite{Lin:2011sa,Gockeler:2011ze,Bratt:2010jn,Yamazaki:2009zq,Yamazaki:2008py}.
(right) The world average of $g_A$ using $N_f=2$ and $2+1$ dynamical lattices results~\cite{Gockeler:2011ze,Pleiter:2011gw,Alexandrou:2010hf,Bratt:2010jn,Yamazaki:2009zq,Lin:2008uz,Lin:2007ap,Khan:2006de}.
}
\end{figure}

The most-calculated nucleon matrix element is the nucleon axial charge, $g_A$. It is one of the most fundamental properties of the nucleon; experimentally, it can be measured though neutron beta decay.
It is also an important parameter that determines the rate of proton-proton fusion, the dominant hydrogen-burning process inside the sun. It can be extracted from the three-point Green function, as shown earlier, but with an axial-vector current inserted:
\begin{equation}
\label{eq:GA}
\langle N| A_\mu| N\rangle = \overline{u_N}(p^\prime) \left[\gamma_\mu\gamma_5 G_A(q^2)+\gamma_5q_\nu \frac{G_P(q^2)}{2M_N} \right]{u_N}(p),
\end{equation}
where ${u_N}$ are nucleon spinors and $q=p^\prime-p$ is the momentum transfer. The nucleon axial charge is defined as $g_A=G_A^{u-d}(q^2=0)$.
The left-hand side of Fig.~\ref{fig:gAT3f} shows a collection of such calculations as a function of quark mass (parametrized by pion mass),
from all $N_f=2+1$ (degenerate up/down and a strange flavors in the sea) lattice calculations.
Most of the results here only display statistical errors, but nonetheless, we can see that the various versions of the Dirac operator generate similar results for this quantity.
Taking an average over all the dynamical results extrapolated to the physical pion mass, we obtain an LQCD $g_A$ of 1.16(3) with about 3\% error, as shown in the right-hand side of Fig.~\ref{fig:gAT3f}.

\subsection{Proton Spin}
\begin{figure}
\begin{center}
\includegraphics[height=.23\textheight]{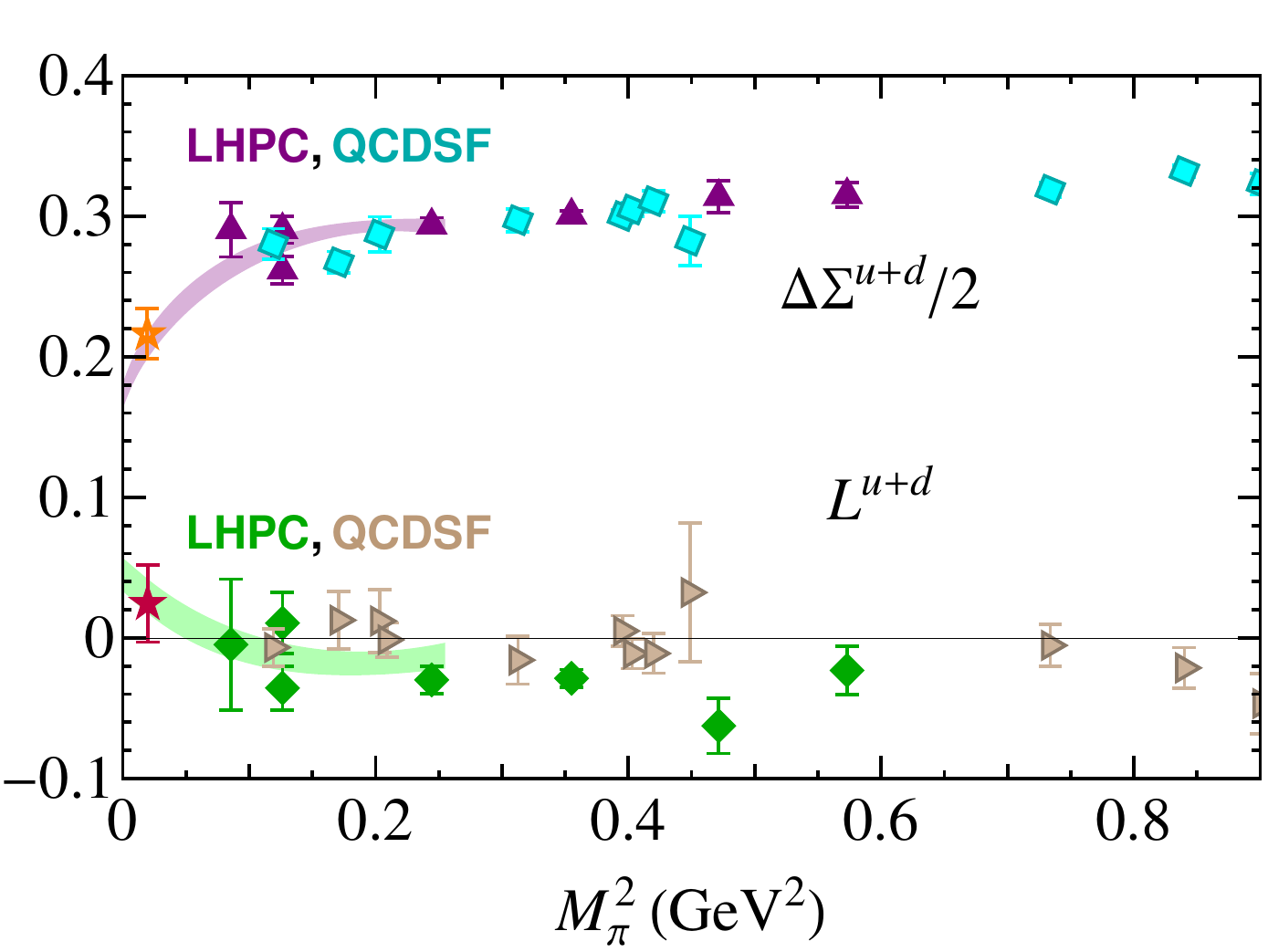}
\includegraphics[height=.23\textheight]{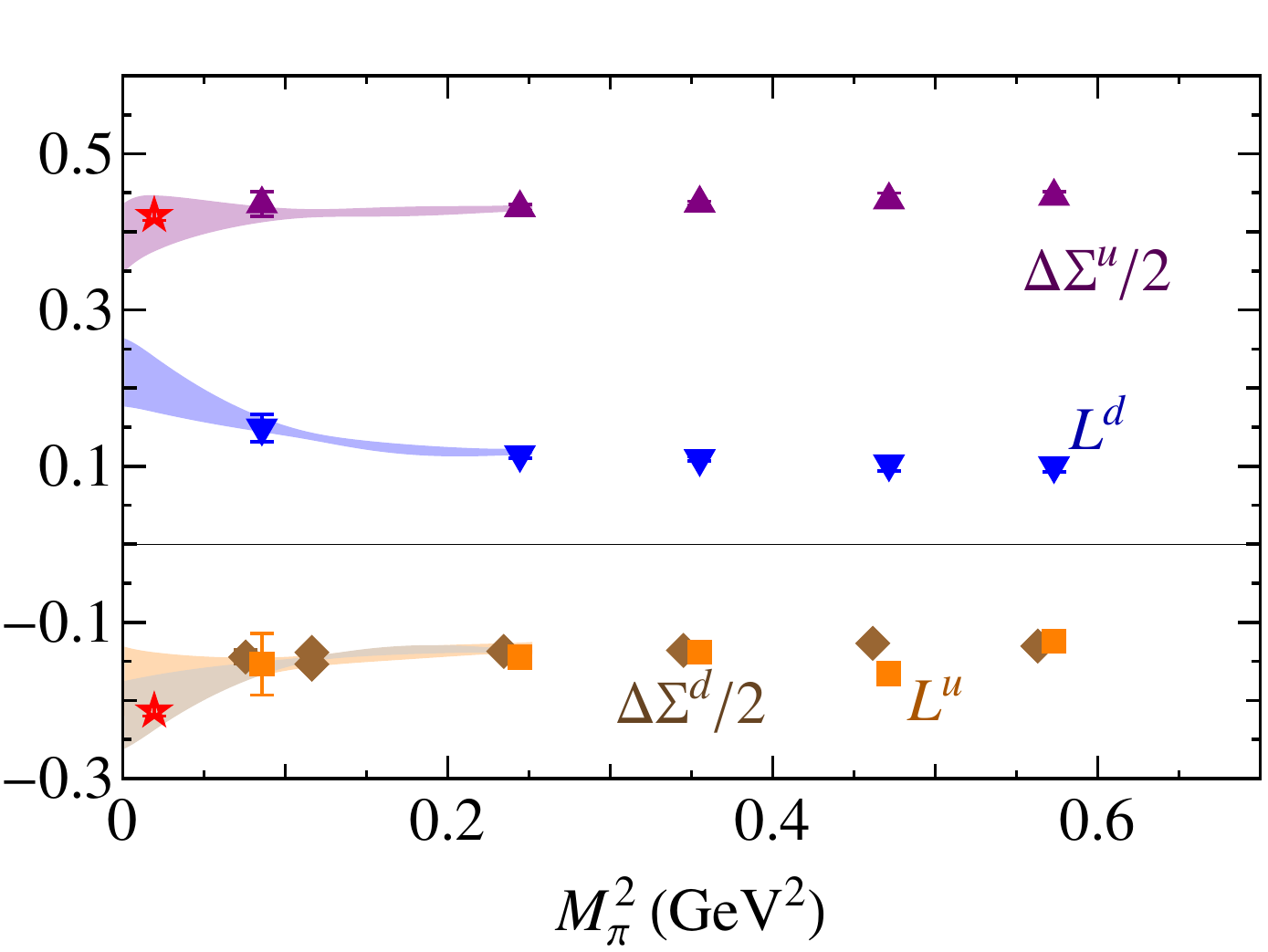}
\end{center}
\caption{\label{fig:spin} (left) Total quark contribution to proton spin and and orbital angular momentum as a function of pion masses by QCDSF and LHPC. The total orbital  angular momentum from quarks is about 6\%.
(right) Decomposition of the individual quark contribution to spin and angular momentum by LHPC. The individual up and down quark contribution to angular momentum is about 36\% in magnitude but opposite in sign; as a result, the total quark contribution is mostly canceled, leaving total around 6\%. The up-quark spin contribution to the total proton spin is about 82\% but the down-quark contributes about half of the size in the opposite sign; leaving total quark spin contribution to proton spin of 40\% or so.
}
\end{figure}

Another fundamental question is how the proton's spin 1/2 is distributed among its constituents.
The most naive intuition is that the three quarks carry the spin. Their contributions can come from their intrinsic spin and orbital angular momentum. We use the set of covariant operators suggested by Ji~\cite{Ji:1995cu} and calculate. Again, we show results as a function of light-quark mass (pion mass).
Figure~\ref{fig:spin} shows a selection of example results calculated by different collaborations  ($N_f=2$ QCDSF~\cite{Brommel:2007sb}
and $N_f=2+1$ LHPC~\cite{Bratt:2010jn}) who use different Dirac operator interpretations (clover and domain-wall fermions on staggered sea, respectively).
These results are renormalized in the $\overline{\rm MS}$ scheme at a scale of 2~GeV.
The extracted values are the stars on the left. As it turns out, the quarks contribute less than 50\% of the total nucleon spin; the majority of it comes from gluons, which is quite a surprise.
However, both of the calculations have ignored the ``disconnected'' diagram; that is, the inserted operator constructed such that its quarks only connect with the valence nucleon quarks through gluons.
Calculating such contributions requires more computational resources to get clean signal, but they could contribute up to 20--30\% of the total quark contributions~\cite{Mathur:1999uf}.
A few groups have made remarkable progress on this type of contribution~\cite{Bali:2011ks,Babich:2010at,Doi:2009sq,Deka:2008xr}, as well as on direct gluon contributions to the spin~\cite{Doi:2008hp}.

\subsection{Transverse Densities}
\begin{figure}
\begin{center}
\includegraphics[width=0.42\columnwidth]{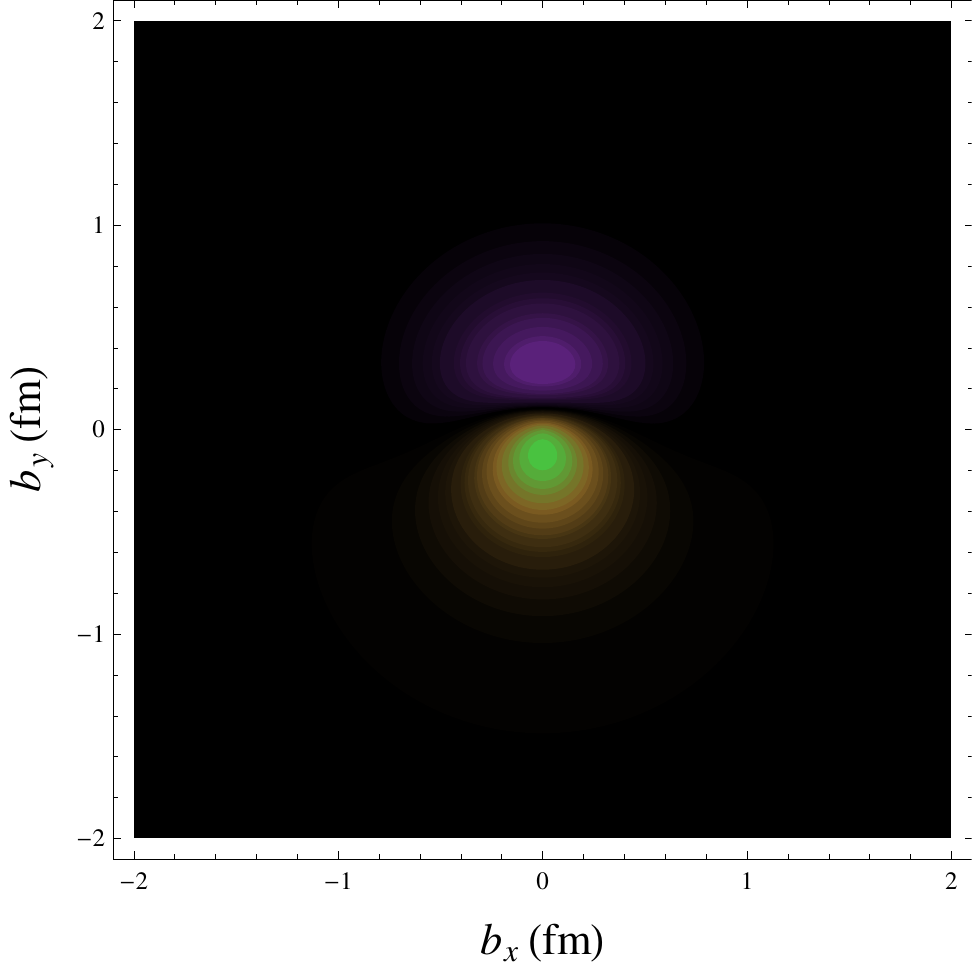}
\includegraphics[width=0.56\columnwidth]{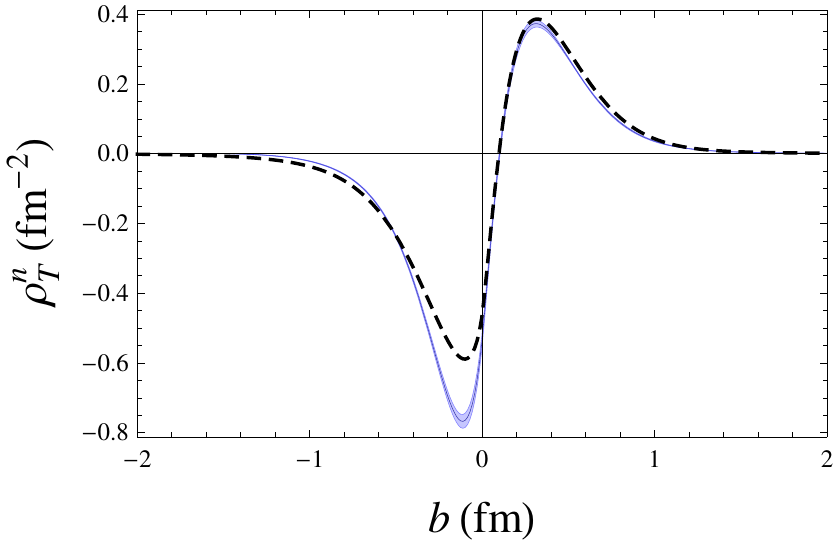}
\end{center}
\caption{The transverse charge densities in a polarized neutron with $b_{x,y}$ (left) and $b$ (right) in fm.
In the left-hand figure, black indicates near-zero values, and purple, orange and white are increasingly positive.
In the right-hand figure, the blue band indicates the densities from 2+1-flavor lattice calculation~\cite{Lin:2010fv}, while the dashed line is a parametrization interpolating and extrapolating from available experimental data~\cite{Arrington:2007ux,Kelly:2004hm}. (Note that the neutron experimental data is extrapolated to larger-$Q^2$ region).
}
\label{fig:F-density}
\end{figure}

Since LQCD gives us access to a complete sample of the QCD vacuum in spacetime, we can also investigate the distribution of charge density within the nucleon.
To avoid relativistic ambiguity due to the transfer of momentum by electromagnetic probes, we should examine the transverse charge density (in directions perpendicular to some impact plane, as an ultrarelativistic electron would see) for a polarized nucleon. Firstly, we need to obtain the nucleon vector-current matrix element over a large range of momenta and from these obtain the electromagnetic form factors ($F_{1,2}(Q^2)$).
Then by taking a sort of Fourier transform, we can see how the transverse charge density is distributed as a function of the impact distance $b$ in a polarized nucleon~\cite{Miller:2010nz,Carlson:2007xd,Miller:2007uy}:
\begin{equation}
\label{eq:rho_T}
\rho_T(b) = \int_0^\infty\!\frac{Q\,dQ}{2\pi}J_0(bQ)F_1(Q^2) + \sin(\phi) \int_0^\infty\!\frac{Q^2\,dQ}{2\pi M_N}J_1(bQ)F_2(Q^2),
\end{equation}
where $J_{0,1}$ are Bessel functions.
We can perform this integral numerically, using the lattice $F_{1,2}(Q^2)$ obtained by extrapolating our fit form to the physical pion mass.
Taking the large-$Q^2$ form-factor data from Ref.~\cite{Lin:2010fv}, Figure~\ref{fig:F-density} shows the results for the neutron in one dimension (right, shown as the blue band here) and the two-dimensional impact plane (left) using our lattice inputs.
There are positive and negative charges surrounding the center, which in the neutron sum to zero.
The dashed line on the left-hand side is a similar application of Eq.~\ref{eq:rho_T} using experimental neutron form factors, which are only known up to $1.5\mbox{ GeV}^2$; 
thus, a majority of the form-factor inputs are based on extrapolation to the larger-$Q^2$ region.
Note that the asymmetry in the distribution for a polarized nucleon is due to the relativistic effect of boosting the magnetic moment of the baryon. This induces an electric dipole moment that shifts the charge distribution.

\section{Applications beyond the Standard Model}

Can we learn about physics beyond the Standard Model (BSM) with LQCD nucleon matrix elements?  There are many opportunities for LQCD to provide important inputs to probe BSM physics with nucleons. For example,
the presence of new particles beyond the Standard Model at the TeV scale can be probed through detection of new scalar and tensor interactions in neutron beta decay. New ultra-cold neutron (UCN) experiments, such as the UCNb/B experiment at Los Alamos National Lab (LANL) or one proposed to run at Oak Ridge National Lab (ORNL) in 2015 can provide high-precision low-energy data to constrain the allowed range of new-particle masses.
The strange contributions to the proton's scalar and spin densities are important inputs for spin-independent and -dependent dark-matter cross-section measurements.

\subsection{Non-$\VmA$ Interactions in Nucleon Beta Decays}

Firstly, we can try to look for non-Standard Model contributions in precision neutron (nuclear) beta-decay measurements.
The neutron beta-decay Hamiltonian contains the Standard-Model $\VmA$ current for the leptons and the quarks, and it may also contain terms corresponding to new BSM physics:
\begin{equation}
H_{\rm eff} = G_F \left( J_{V-A}^{\rm lept} \times J_{V-A}^{\rm quark} + \sum_i \varepsilon_i^{\rm BSM} \hat{O}_i^{\rm lept} \times \hat{O}_i^{\rm quark}, \right),
\end{equation}
where $G_F$ is the Fermi constant, $J_{V-A}$ is the left-handed current of the indicated particle, and the sum includes operators with novel chiral structure.
So in the context of our theory, new operators will enter with the coefficients $\varepsilon$ that are related to the TeV scale of the particles. The leptonic part is understandable using analytic techniques, but the quark operator in the context of the nucleon will introduce some unknown coupling constants:
\begin{equation}
g_T = \langle n | \overline{u}\sigma_{\mu\nu} d | p \rangle,\; g_S = \langle n | \overline{u} d | p \rangle,
\end{equation}
which are nonperturbative functions of the nucleon structure, described in the SM by QCD.
Any deviation from the SM $\VmA$ current coming from new scalar and tensor interactions in the effective theory will require knowledge of the couplings $g_S$ and $g_T$ to understand.
For more details about experimental and theoretical work on this subject, we refer readers to Ref.~\cite{Bhattacharya:2011qm}, and references within.

\begin{figure}
\begin{center}
\includegraphics[width=.48\textwidth]{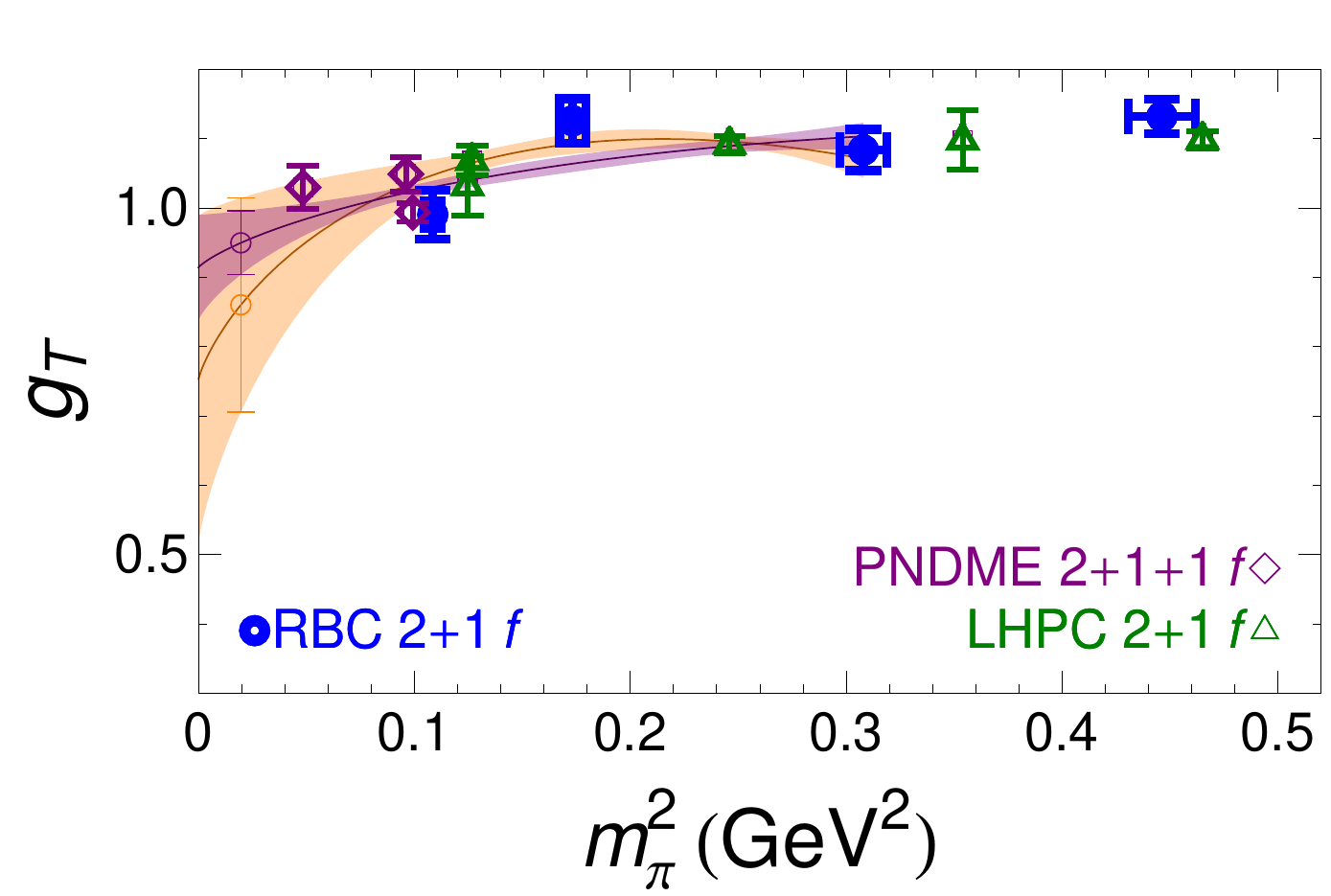}
\includegraphics[width=.48\textwidth]{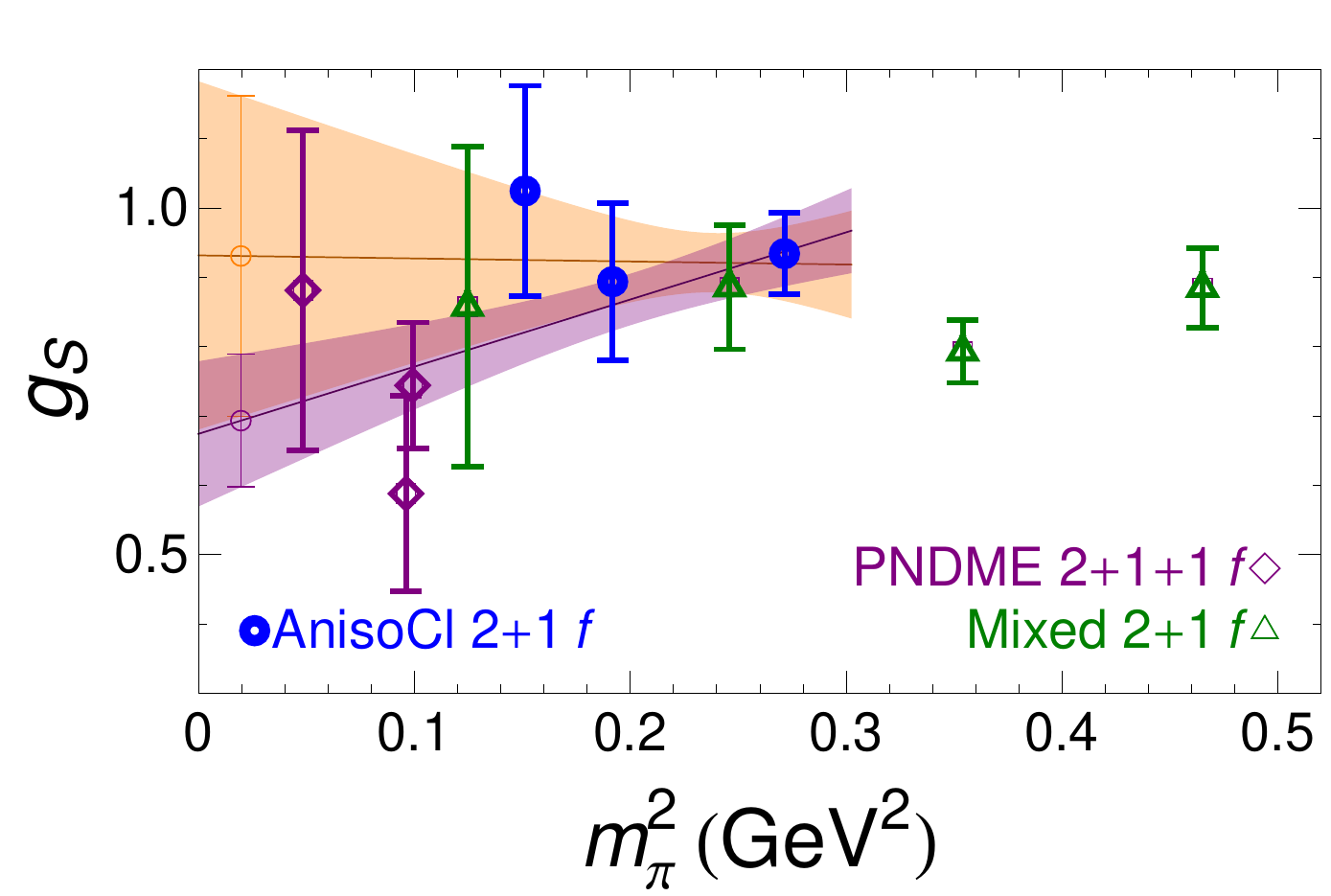}
\end{center}
\caption{\label{fig:gTS}
Global analysis of all $N_f=2+1$ lattice calculations of $g_T$ (left) and $g_S$ (right). The leftmost points are the extrapolated values at the physical pion mass. The two bands show different extrapolations in the pion mass: with and without the lighter PNDME collaboration points from Ref.~\cite{Lin:2011xx}; the final uncertainty due to the chiral extrapolation is significantly improved.
}
\end{figure}

The tensor charge  $g_T$  has been studied a few times in the past using $N_f=2+1$ dynamical ensembles in lattice QCD, as summarized in the left-hand side of Fig.~\ref{fig:gTS}.
We take the 2+1-flavor results by the RBC~\cite{Aoki:2010xg} and LHPC~\cite{Edwards:2006qx}, and recently by PNDME~\cite{Lin:2011xx}, and make a global plot as a function of pion mass squared. $g_S$, on the other hand, has not been much studied; a few collected calculations~\cite{Lin:2011xx} is shown on the right-hand side of Fig.~\ref{fig:gTS} using various $N_f=2+1$ lattice actions.
All of these results, by different collaborations, using different fermion actions, (but at the same lattice spacing) are generally in good agreement, although the errorbars shown here contain only statistical error.
We further globally analyze all the available lattice data (with a upper pion-mass cut at 550~MeV), including (excluding) the data from PNDME (whose lightest pion mass is around 220~MeV), shown as the purple (orange) band in Fig.~\ref{fig:gTS}. We use the chiral formulation given in Ref.~\cite{Detmold:2002nf} and a linear ansatz for tensor and scalar charges, respectively, to extrapolate to the physical pion mass. We see that the PNDME points greatly constrain the uncertainty due to chiral extrapolation in both cases and obtain $g_T^{\rm LQCD}= 0.95(5)$ and $g_S^{\rm LQCD}= 0.69(9)$. 
The LQCD values are better determined than other theoretical estimations (from different model approximations), which give rather loose bounds on these quantities; for example, $g_S$ is estimated to be between 0.25 and 1.

\begin{figure}
\begin{center}
\includegraphics[width=.44\textwidth]{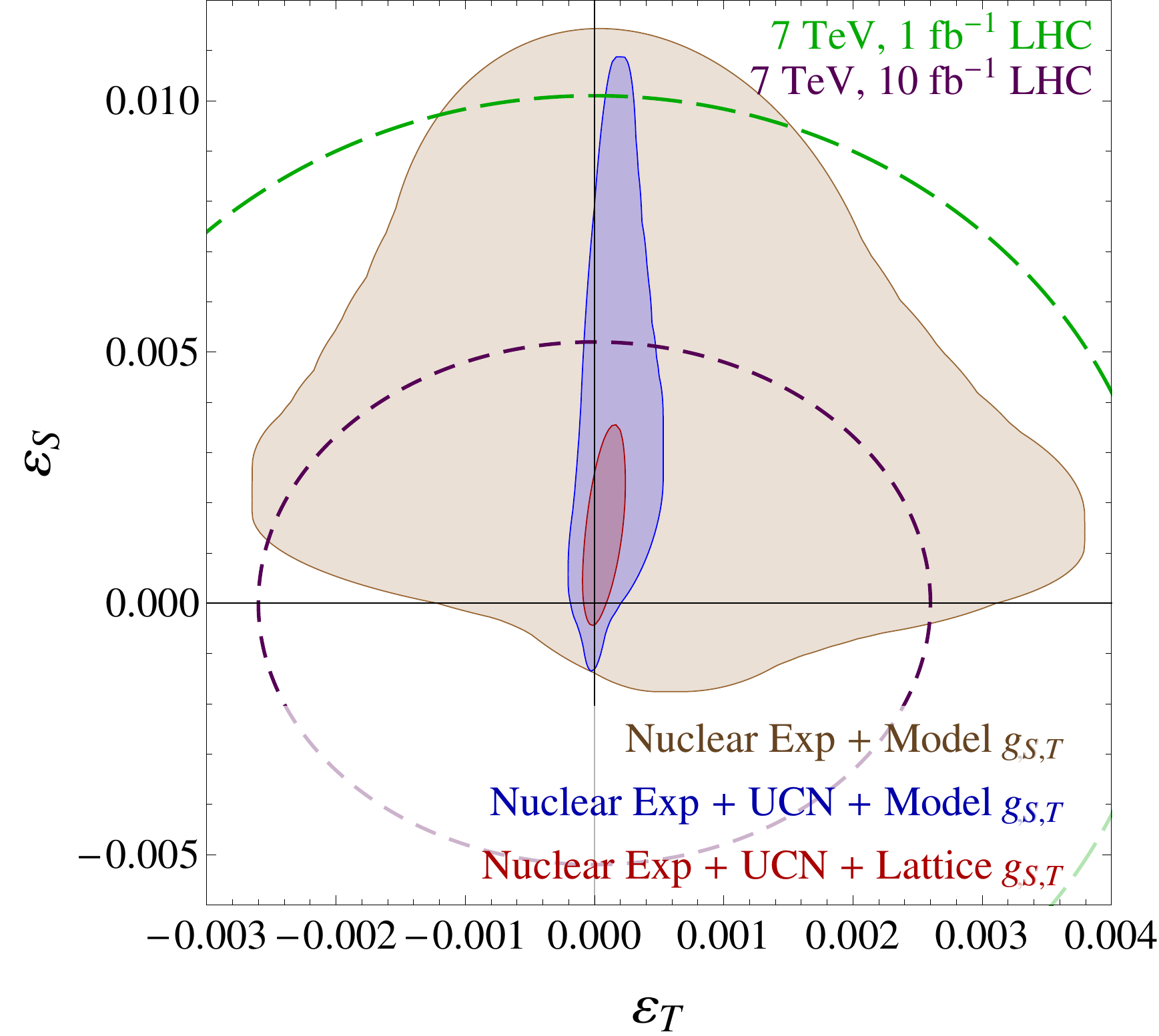}
\includegraphics[width=.54\textwidth]{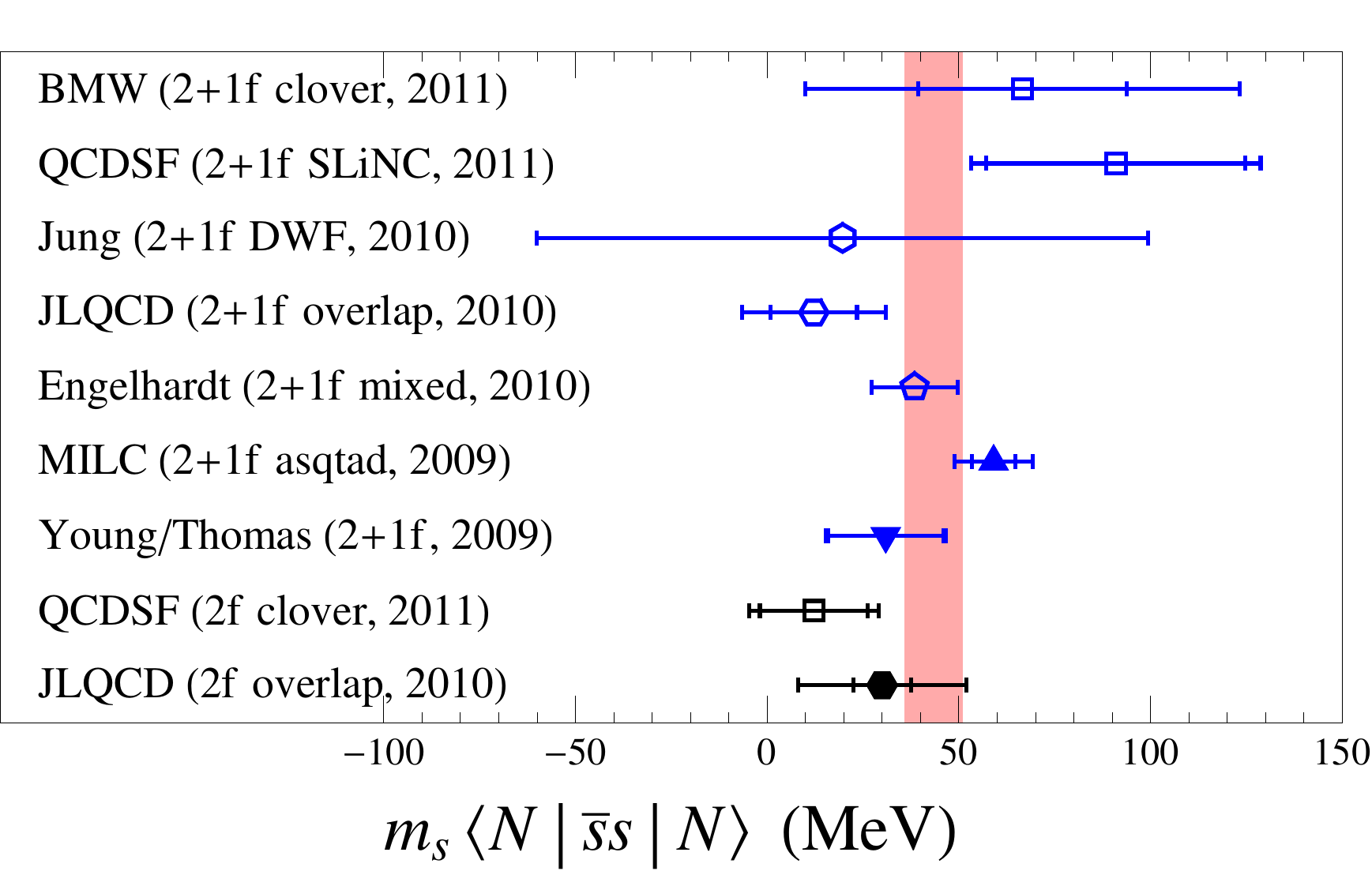}
\end{center}
\caption{\label{fig:fig3}
(left) $\epsilon_{S}$-$\epsilon_{T}$ allowed parameter region using different experimental and theoretical inputs. The green and purple dashed lines are the constraints from LHC current bounds and near-term expectation.
(right) Summary of dynamical LQCD calculations of the nucleon sigma term $m_s\langle \overline{s}s\rangle$ in MeV.
The (pink) band indicates the global average after weighting individual points according to their systematics.
}
\end{figure}

Finally, these low-energy matrix elements, the tensor and scalar charges, can be combined with experimental data to determine the allowed region of parameter space for scalar and tensor BSM couplings (denoted $\epsilon$).
Using the $g_{S,T}$ from the model estimations and combining with the existing nuclear experimental data\footnote{For example, nuclear beta decay $0^+ \rightarrow 0^+$ transitions and other processes, such as $\beta$ asymmetry in Gamow-Teller $^{60}\mbox{Co}$,
longitudinal polarization ratio between Fermi and Gamow-Teller transitions in $^{114}\mbox{In}$,
positron polarization in polarized $^{107}\mbox{In}$
and beta-neutrino correlation parameters in nuclear transitions.}, we get the constraints shown as the outermost band on the left-hand side of Fig.~\ref{fig:fig3}.
An ongoing UCN experiment is studying neutron beta decay at LANL to look for deviations from the SM in
the Fierz term and the neutrino asymmetry parameter 
to the level of $10^{-3}$ by 2013. Combining those expected data and existing measurements, and again, using the model inputs of $g_{S,T}$, we see the uncertainties in $\epsilon_{S,T}$ are significantly improved.
Finally, using our present lattice-QCD values of the scalar and tensor charges, combined with the expected 2013 precision of experimental bounds on deviation of these neutron-decay parameters from their SM values,
we found the constraints on $\epsilon_{S,T}$ are further improved, shown as the innermost region.
These upper bounds on the effective couplings $\epsilon_{S,T}$ correspond to lower bounds for the scales $\Lambda_{S,T}$ at 2.9 and 5.6~TeV, 
respectively, for new physics in these channels.

How do the constraints from high-energy experiments compare? As demonstrated in Ref.~\cite{Bhattacharya:2011qm}, neither CDF nor D0 is sufficient to provide useful constraints.
We can estimate the $\epsilon_{S,T}$ constraints from LHC current bounds and near-term expectations through effective Lagrangian
\begin{equation}
{\cal L} = -\frac{\eta_S}{\Lambda_S^2}V_{\rm ud}(\overline{u}d)(\overline{e}P_L\nu_e)-\frac{\eta_T}{\Lambda_T^2}V_{\rm ud}(\overline{u}\sigma^{\mu\nu}d)(\overline{e}\sigma_{\mu\nu}P_L\nu_e),
\end{equation}
where $\eta_{S,T}=\pm 1$. 
By looking at events with high transverse mass from CMS/ATLAS in the $e\nu+X$ channel and comparing with the SM $W$ background, we estimated 90\%-C.L. constraints on $\eta_{S,T}$ based on the current LHC run, $\sqrt{s}=7$~TeV ${\cal L}=1\mbox{ fb}^{-1}$ (the green line) and for a near-future run $\sqrt{s}=7$~TeV ${\cal L}=10\mbox{ fb}^{-1}$ (the purple dashed line) on the left-hand side of Fig.~\ref{fig:fig3}. More details can be found in Ref.~\cite{Bhattacharya:2011qm}.

\subsection{Dark Matter Cross-Sections}

Recent evidence suggests that the composition of dark matter must be cold, with velocities that are non-relativistic, giving it a Compton wavelength above the width of a proton. One promising candidate for dark matter is the supersymmetric particle called the neutralino. If dark matter is neutralinos, they will interact with nuclei at low-energy. A typical value for the cross-section in the Constrained Minimal Supersymmetric Standard Model (CMSSM) is about 1 zeptobarn ($10^{-49}\mbox{ m}^2$).
The strange-quark contribution to the proton scalar density and spin are important inputs for spin-independent and -dependent cross-sections. In this proceeding, we will focus on the LQCD calculation of $\sigma_s=m_s\langle N|\bar{s}s| N\rangle$.

Many groups have developed techniques to investigate $\sigma_s$.
One can directly calculate the matrix element which involves strange-quark loops and proton correlators. Unfortunately, such a direct calculation is noisy, so various techniques have been developed to improve the signal (e.g. JLQCD, QCDSF, BU, Engelhardt)~\cite{Ohki:2008ff,Bali:2009dz,Babich:2010at,Takeda:2010cw,Collins:2010gr,:2010id,Engelhardt:2010zr}. Alternatively, one could use the Feynman-Hellman Theorem by studying how the nucleon mass varies with the strange-quark mass, taking the numerical derivative $dM_N/dm_s$ either by direct SU(3) fitting to the baryon masses (e.g. Young/Thomas) or by reweighting the strange part of action (e.g. Jung, MILC)~\cite{Toussaint:2009pz,Young:2009zb,Jung:2010}.
The values of $\sigma_s$ from various calculations are shown on the left-hand side of Fig.~\ref{fig:fig3}.
We perform a global fit to current dynamical lattice data (weighted by lattice spacing, lightest $m_\pi$, dynamical strange and other quality factors) and obtain $\sigma_s^{\rm LQCD}\!\!\!=43(8)$~MeV.
The value can be input into theoretical calculations of cross-section constraints from various models; an example of the application to the CMSSM can be found in Ref.~\cite{Giedt:2009mr}.


\section{Road to High Precision}

Nucleon matrix elements are generally more computationally demanding than mesonic matrix elements, since the statistical noise grows with Euclidean time $t$ as $e^{(M_N-3M_\pi/2)t}$ for each nucleon in the system.
Such characteristic signal-to-noise behavior has been studied and well-demonstrated up to small nuclei using high statistics by NPLQCD~\cite{Beane:2011pc,Beane:2009py,Beane:2009gs,Beane:2009kya}.
Take $g_A$, for example; the present uncertainty (including the systematics) is around 6--10\% \cite{Bratt:2010jn,Yamazaki:2008py,Lin:2008uz,Lin:2007ap}.
If we are to target statistical error for $g_A$ at a level of 1\% or smaller, we need at least an order-of-magnitude boost in computational resources.

Anther obstacle to obtaining high-precision nucleon matrix elements is to have control over the systematic uncertainties of the LQCD calculation. The lattice scales, the lattice spacing and box size, must be taken $a\rightarrow 0$ and $L \rightarrow \infty$, or we must at least show that the quantities of interest are independent of these unphysical scales.
The nearest excited state of the nucleon, the Roper resonance, is only few hundred MeV away, unlike the $\pi^\prime$, which is almost an order of magnitude heavier than the pion. As a result, it can potentially create significant contamination to the wanted nucleon matrix element. One way to control this systematic is to increase the source-sink separation in the matrix-element calculation, since the  excited-state contamination is exponentially suppressed; however, this will also decrease the signal-to-noise ratio of the desired matrix element as the separation enlarges.
Refs.~\cite{Lin:2011sa,Lin:2010fv} have explicitly taken account of the excited-state contribution to the nucleon matrix element in their analysis.

Chiral extrapolation of nucleon quantities can also introduce large systematics, especially when the formulation contains significant contributions from chiral logarithms. Figure~\ref{fig:gTS} clearly demonstrates how an extrapolation could be improved by the inclusion of calculations at lighter pion mass. Such extrapolations also raise the questions of the validity of chiral perturbation theory over the range of pion masses used, how a particular expansion scale adopted converged, and whether one should use expansions with SU(3) or SU(2) flavor symmetry. In all cases, having light pion masses is essential and many collaborations already have begun generating gauge ensembles at the physical pion mass, which will soon remove extrapolation systematics.

To compare lattice results with continuum ones, we also need to renormalize the lattice operators used in Eq.~\ref{eq:threepoint} and match to the continuum $\overline{\mbox{MS}}$ scheme (running to 2~GeV scale). Unlike the continuum case, calculating the lattice-operator renormalization factor in perturbation theory is complicated and slowly converging. To reduce the systematic associated with renormalization, it is commonly agreed that using a nonperturbative approach (such as Schr\"odinger Functional or RI/MOM scheme) on the lattice should work best. To extract high-precision nucleon matrix elements, we need to improve the accuracy of nonperturbative renormalization as well as working out the continuum matching to several-loop order for each operator of interest.

With the ever increasing computational resources made available to us, we can overcome the challenges that occur in nucleon systems to improve the uncertainties on nucleon matrix elements.
Furthermore, extrapolation to the physical light-quark masses is likely to be brought under control in the next few years, as ensembles of lattices begin to be generated with physical $u$, $d$ (and $s$ and $c$) quark masses, which should greatly reduce the systematic uncertainties.
Other systematics, such as finite-volume effects, renormalization and excited-state contamination can be reduced by improved algorithms and by increasing the computational resources devoted to the calculations.

To sum up, the name of the game is precision, and lattice QCD is now entering an exciting era where we can provide precision input directly from the first principles of the Standard Model.
This is partially thanks to the global increase in computational resources available to us, and also due to improved algorithms, which have greatly improved our abilities to calculate quantities that were not accessible before.
There are many quantities that we can contribute using supercomputers.
We see various different lattice actions and groups with independent calculations provide consistency checks. We have reproduced well measured experimental values and provided better SM values. We can make predictions for quantities that have not or could not be measured by experiment. Lattice QCD is essential in cases where experiments are limited, such as for the scalar charge or hyperon axial couplings.

New experiments at both the high-energy frontier and the precision low-energy frontier are now enabling us to probe BSM physics. But to do so, we need precision quantities from Standard Model, such as $g_{S,T}$ or the strangeness in proton, to constrain various possible BSM models.
By continuing to improve the precision of our nucleon matrix elements, we can play a considerable role in this search. We will also be able to identify high-impact contributions that lattice techniques can make. These could be for quantities which are difficult or impossible to measure in experiment and where models introduce unacceptable levels of uncertainty. By determining what needs to be computed and delivering precise values from first principles, lattice QCD will solidify its role in particle and nuclear physics.

\section*{Acknowledgments}
The speaker is supported by DOE grant DE-FG02-97ER4014.



\section*{References}
\begin{thebibliography}{10}

\bibitem{Joo:2011xx}
Balint Joo.
\newblock {Challenges and Opportunities for Lattice QCD Software}.
\newblock {\em this conference procceding}, 2011.

\bibitem{Lin:2011sa}
Huey-Wen Lin and Saul~D. Cohen.
\newblock {Nucleon and Pion Form Factors from $N_f=2+1$ Anisotropic Lattices}.
\newblock 2011.

\bibitem{Gockeler:2011ze}
M.~Gockeler et~al.
\newblock {Baryon Axial Charges and Momentum Fractions with $N_f=2+1$ Dynamical
  Fermions}.
\newblock {\em PoS}, LATTICE2010:163, 2010.

\bibitem{Bratt:2010jn}
J.D. Bratt et~al.
\newblock {Nucleon structure from mixed action calculations using 2+1 flavors
  of asqtad sea and domain wall valence fermions}.
\newblock {\em Phys. Rev.}, D82:094502, 2010.

\bibitem{Yamazaki:2009zq}
Takeshi Yamazaki et~al.
\newblock {Nucleon form factors with 2+1 flavor dynamical domain-wall
  fermions}.
\newblock {\em Phys. Rev.}, D79:114505, 2009.

\bibitem{Yamazaki:2008py}
T.~Yamazaki et~al.
\newblock {Nucleon axial charge in 2+1 flavor dynamical lattice QCD with domain
  wall fermions}.
\newblock {\em Phys. Rev. Lett.}, 100:171602, 2008.

\bibitem{Pleiter:2011gw}
D.~Pleiter et~al.
\newblock {Nucleon form factors and structure functions from N(f)=2 Clover
  fermions}.
\newblock {\em PoS}, LATTICE2010:153, 2010.

\bibitem{Alexandrou:2010hf}
C.~Alexandrou et~al.
\newblock {Axial Nucleon form factors from lattice QCD}.
\newblock {\em Phys. Rev.}, D83:045010, 2011.

\bibitem{Lin:2008uz}
Huey-Wen Lin, Tom Blum, Shigemi Ohta, Shoichi Sasaki, and Takeshi Yamazaki.
\newblock {Nucleon structure with two flavors of dynamical domain- wall
  fermions}.
\newblock {\em Phys. Rev.}, D78:014505, 2008.

\bibitem{Lin:2007ap}
Huey-Wen Lin and Konstantinos Orginos.
\newblock {First Calculation of Hyperon Axial Couplings from Lattice QCD}.
\newblock {\em Phys. Rev.}, D79:034507, 2009.

\bibitem{Khan:2006de}
A.~Ali Khan et~al.
\newblock {Axial coupling constant of the nucleon for two flavours of dynamical
  quarks in finite and infinite volume}.
\newblock {\em Phys. Rev.}, D74:094508, 2006.

\bibitem{Ji:1995cu}
Xiang-Dong Ji, Jian Tang, and Pervez Hoodbhoy.
\newblock {The spin structure of the nucleon in the asymptotic limit}.
\newblock {\em Phys. Rev. Lett.}, 76:740--743, 1996.

\bibitem{Brommel:2007sb}
Dirk Brommel et~al.
\newblock {Moments of generalized parton distributions and quark angular
  momentum of the nucleon}.
\newblock {\em PoS}, LAT2007:158, 2007.

\bibitem{Mathur:1999uf}
N.~Mathur, S.~J. Dong, K.~F. Liu, L.~Mankiewicz, and N.~C. Mukhopadhyay.
\newblock {Quark orbital angular momentum from lattice QCD}.
\newblock {\em Phys. Rev.}, D62:114504, 2000.

\bibitem{Bali:2011ks}
Gunnar~S. Bali et~al.
\newblock {The strange and light quark contributions to the nucleon mass from
  Lattice QCD}.
\newblock 2011.

\bibitem{Babich:2010at}
Ronald Babich et~al.
\newblock {Exploring strange nucleon form factors on the lattice}.
\newblock 2010.

\bibitem{Doi:2009sq}
Takumi Doi, Mridupawan Deka, Shao-Jing Dong, Terrence Draper, Keh-Fei Liu,
  et~al.
\newblock {Nucleon strangeness form factors from N(f) = 2+1 clover fermion
  lattice QCD}.
\newblock {\em Phys. Rev.}, D80:094503, 2009.

\bibitem{Deka:2008xr}
M.~Deka, T.~Streuer, T.~Doi, S.J. Dong, T.~Draper, et~al.
\newblock {Moments of Nucleon's Parton Distribution for the Sea and Valence
  Quarks from Lattice QCD}.
\newblock {\em Phys. Rev.}, D79:094502, 2009.

\bibitem{Doi:2008hp}
Takumi Doi et~al.
\newblock {Strangeness and glue in the nucleon from lattice QCD}.
\newblock {\em PoS}, LATTICE2008:163, 2008.

\bibitem{Miller:2010nz}
Gerald~A. Miller.
\newblock {Transverse Charge Densities}.
\newblock {\em Ann. Rev. Nucl. Part. Sci.}, 60:1--25, 2010.

\bibitem{Carlson:2007xd}
Carl~E. Carlson and Marc Vanderhaeghen.
\newblock {Empirical transverse charge densities in the nucleon and the
  nucleon-to-$\Delta$ transition}.
\newblock {\em Phys. Rev. Lett.}, 100:032004, 2008.

\bibitem{Miller:2007uy}
Gerald~A. Miller.
\newblock {Charge Density of the Neutron}.
\newblock {\em Phys. Rev. Lett.}, 99:112001, 2007.

\bibitem{Lin:2010fv}
Huey-Wen Lin, Saul~D. Cohen, Robert~G. Edwards, Kostas Orginos, and David~G.
  Richards.
\newblock {Lattice Calculations of Nucleon Electromagnetic Form Factors at
  Large Momentum Transfer}.
\newblock 2010.

\bibitem{Arrington:2007ux}
J.~Arrington, W.~Melnitchouk, and J.~A. Tjon.
\newblock {Global analysis of proton elastic form factor data with two-photon
  exchange corrections}.
\newblock {\em Phys. Rev.}, C76:035205, 2007.

\bibitem{Kelly:2004hm}
J.~J. Kelly.
\newblock {Simple parametrization of nucleon form factors}.
\newblock {\em Phys. Rev.}, C70:068202, 2004.

\bibitem{Bhattacharya:2011qm}
Tanmoy Bhattacharya et~al.
\newblock {Probing Novel Scalar and Tensor Interactions from (Ultra)Cold
  Neutrons to the LHC}.
\newblock 2011.

\bibitem{Aoki:2010xg}
Yasumichi Aoki, Tom Blum, Huey-Wen Lin, Shigemi Ohta, Shoichi Sasaki, et~al.
\newblock {Nucleon isovector structure functions in (2+1)-flavor QCD with
  domain wall fermions}.
\newblock {\em Phys. Rev.}, D82:014501, 2010.

\bibitem{Edwards:2006qx}
R.~G. Edwards et~al.
\newblock {Nucleon structure in the chiral regime with domain wall fermions on
  an improved staggered sea}.
\newblock {\em PoS}, LAT2006:121, 2006.

\bibitem{Lin:2011xx}
Huey-Wen Lin et~al.
\newblock {Probing TeV Physics through Neutron-Decay Matrix Elements}.
\newblock 2011.

\bibitem{Detmold:2002nf}
William Detmold, W.~Melnitchouk, and Anthony~William Thomas.
\newblock {Moments of isovector quark distributions from lattice QCD}.
\newblock {\em Phys. Rev.}, D66:054501, 2002.

\bibitem{Ohki:2008ff}
H.~Ohki et~al.
\newblock {Nucleon sigma term and strange quark content from lattice QCD with
  exact chiral symmetry}.
\newblock {\em Phys. Rev.}, D78:054502, 2008.

\bibitem{Bali:2009dz}
Gunnar Bali, Sara Collins, and Andreas Schafer.
\newblock {Strangeness and charm content of the nucleon}.
\newblock {\em PoS}, LAT2009:149, 2009.

\bibitem{Takeda:2010cw}
K.~Takeda et~al.
\newblock {Nucleon strange quark content from two-flavor lattice QCD with exact
  chiral symmetry}.
\newblock {\em Phys. Rev.}, D83:114506, 2011.

\bibitem{Collins:2010gr}
Sara Collins et~al.
\newblock {Disconnected contributions to hadronic structure}.
\newblock {\em PoS}, LATTICE2010:134, 2010.

\bibitem{:2010id}
K.~Takeda et~al.
\newblock {Nucleon strange quark content in 2+1-flavor QCD}.
\newblock {\em PoS}, LATTICE2010:160, 2010.

\bibitem{Engelhardt:2010zr}
Michael Engelhardt.
\newblock {Strangeness in the nucleon from a mixed action calculation}.
\newblock {\em PoS}, LATTICE2010:137, 2010.

\bibitem{Toussaint:2009pz}
D.~Toussaint and W.~Freeman.
\newblock {The strange quark condensate in the nucleon in 2+1 flavor QCD}.
\newblock {\em Phys. Rev. Lett.}, 103:122002, 2009.

\bibitem{Young:2009zb}
R.~D. Young and A.~W. Thomas.
\newblock {Octet baryon masses and sigma terms from an SU(3) chiral
  extrapolation}.
\newblock {\em Phys. Rev.}, D81:014503, 2010.

\bibitem{Jung:2010}
Chulwoo Jung.
\newblock {\em LAT2010}, 2010.

\bibitem{Giedt:2009mr}
Joel Giedt, Anthony~W. Thomas, and Ross~D. Young.
\newblock {Dark matter, the CMSSM and lattice QCD}.
\newblock {\em Phys. Rev. Lett.}, 103:201802, 2009.

\bibitem{Beane:2011pc}
S.~R. Beane et~al.
\newblock {High Statistics Analysis using Anisotropic Clover Lattices: (IV)
  Volume Dependence of Light Hadron Masses}.
\newblock {\em Phys. Rev.}, D84:014507, 2011.

\bibitem{Beane:2009py}
Silas~R. Beane et~al.
\newblock {High Statistics Analysis using Anisotropic Clover Lattices: (III)
  Baryon-Baryon Interactions}.
\newblock {\em Phys. Rev.}, D81:054505, 2010.

\bibitem{Beane:2009gs}
Silas~R. Beane et~al.
\newblock {High Statistics Analysis using Anisotropic Clover Lattices: (II)
  Three-Baryon Systems}.
\newblock {\em Phys. Rev.}, D80:074501, 2009.

\bibitem{Beane:2009kya}
Silas~R. Beane et~al.
\newblock {High Statistics Analysis using Anisotropic Clover Lattices: (I)
  Single Hadron Correlation Functions}.
\newblock {\em Phys. Rev.}, D79:114502, 2009.

\end{thebibliography}
\end{document}